\documentclass{article}
\usepackage[utf8]{inputenc}
\usepackage{authblk}
\usepackage[english]{babel}

\usepackage[a4paper,top=2cm,bottom=2cm,left=3cm,right=3cm,marginparwidth=1.75cm]{geometry}

\usepackage{amsmath}
\usepackage{graphicx}
\usepackage[colorlinks=true, allcolors=blue]{hyperref}
\def\simpropto{\lower.2ex\hbox{$\; \buildrel \propto \over \sim \;$}}
\def\ltsim{\lower.5ex\hbox{$\; \buildrel < \over \sim \;$}}
\def\gtrsim{\lower.5ex\hbox{$\; \buildrel > \over \sim \;$}}

\graphicspath{{Figures/}}

\title{Lensing Constraints on PBHs: Substellar to Intermediate Masses}

\author[1,2]{\bf Evencio Mediavilla}
\author[3,4]{\bf Jorge Jim\'enez-Vicente}
\affil[1]{Instituto de Astrofísica de Canarias, Vía L\'actea S/N, E-38200 La Laguna, Tenerife, Spain}
\affil[2]{Departamento de Astrofísica, Universidad de la Laguna, E-38200 La Laguna, Tenerife, Spain}
\affil[3]{Departamento de Física Te\'orica y del Cosmos, Universidad de Granada, Campus de Fuentenueva, E-18071 Granada, Spain}
\affil[4]{Instituto Carlos I de Física Te\'orica y Computacional, Universidad de Granada, E-18071 Granada, Spain }

\begin{document}
\maketitle\label{chapter:lensing}

\begin{abstract}
 \noindent

 Gravitational microlensing is a robust tool to detect and directly measure the abundance and mass of any kind of compact objects, either in our galaxy or in the extragalatic domain. On basis to generic, broadly applicable arguments, it is concluded that the observed microlensing magnifications are too small and the microlensing events less frequent than the expectations for a significant population of compact objects (other than normal stars). The detection of chromatic effects of microlensing, neither supports the presence of BHs. Detailed statistical studies of the observed microlensing magnifications and events frequency impose strict upper limits to the fraction of total mass of BHs ($\ltsim$ 1\%) from $10^{-7}M_\odot$ to indefinitely large masses. These results hold even when the BHs are distributed according to a mass spectrum or are forming clusters.
\end{abstract}

\section{Lensing effects of compact objects}

Since the experimental confirmation of Einstein’s principle of equivalence and general theory of relativity,  we know that massive objects, no matter their origin or nature, deflect light rays. This property is used to measure mass in different astrophysical and cosmological scenarios using the Gravitational Lensing theory.

Regarding the study of PBHs we are particularly interested in the detection of the possible gravitational lensing effects induced by the presence of these objects in galaxies and clusters of galaxies. Specifically, this chapter focuses on the consequences of lensing studies in our galaxy and, especially, in lensed quasars. Lensing allows to detect and study the presence of compact massive objects in at least two ways: globally (through macro-modeling anomalies) and locally (from microlensing effects). Macro-modeling fits the positions of the multiple images of a lensed quasar, using a suitable smooth mass distribution of the lens galaxy and predicts, among other observables, the flux-ratios between images. Consequently, positional (astrometric) and flux-ratio anomalies (greater than model allowances arising from experimental uncertainties) may indicate the presence of massive objects not considered in the modeling process. Microlensing effects\footnote{Also millilensing effects (micro and millilensing are commonly distinguished by the strength of the effect).} are caused by the granulation of the lens galaxy mass distribution in compact objects and mainly consist in the change of flux of any of the source images caused by the presence of these objects along the observer’s line of sight. Thus, while  microlensing is a local effect directly related to the presence of compact objects, macro-modeling anomalies can be affected by degeneracies in the description of the global mass distribution of the lens galaxy and are, consequently, more model dependent. For this reason, micro and millilensing are especially interesting in the context of PBHs although we also shall consider some inputs from macro-modeling.

From the experimental point of view, there are two approaches commonly used to study microlensing: monitoring of flux variability of lensed quasar images (light-curve method) and comparison of the continuum flux-ratio between two images with respect to their intrinsic flux-ratio, either determined from the flux of a very large region insensitive to microlensing or from a model (single-epoch method). In practical time intervals,  millilensing can be studied only using the second method. 

Two general properties of gravitational lensing are worthy to be highlighted for they can help to discuss the existence of massive compact objects. The first one is mass-length degeneracy (a result of the scale invariance of gravitation, Schechter et al. 2014, Esteban-Gutierrez et al. 2020), which has the consequence that for a source of fixed size and a population of compact objects of a given mass density, the larger the mass of the compact objects the greater the impact of microlensing. The second one is the dependence of microlensing strength with the source size (microlensing chromaticity). In principle, microlensing is sensitive to the source size relative to the Einstein radius\footnote{The Einstein radius, used as the natural scale of microlensing, increases with the square root of the microlens mass and corresponds to the radius of the region in which the magnification induced by an isolated microlens is noticeable.} of the compact object: the smaller the source, the larger the strength of microlensing. However, in terms of the Einstein radius of the microlens, there is a low limit of the source size (point source limit) below which microlensing progressively approaches a constant value and no significant chromaticity is further present. As we comment along the chapter, none of these two properties support the existence of PBHs.

This chapter is organized as follows. For historical reasons and to introduce the main concepts, we start with galactic MACHOS studies in the Milky Way, Andromeda and other nearby galaxies in Section \ref{GalMACHO} and with the early work on the extension of the MACHOS experimental idea to the extragalactic domain in Section \ref{ExGalMACHO} (mainly focused on typical stellar masses). Above the stellar mass range, we  study the very interesting (after LIGO-Virgo discoveries) case of the intermediate mass PBHs in Section \ref{ExGalMACHO:LIGO}. Below the stellar mass-range, in Section \ref{ExGalMACHO:substel} we consider the sub-stellar domain from X-ray observations. In Section \ref{massspecandclus} we abandon the simplification of an spatially uniform distribution of single mass microlenses, to explore both, the effect of a spectrum of masses and the impact of clustering (this last study can be extrapolated to indefinitely large masses). The chapter ends with a section devoted to summarize the main results.


\section{Galactic MACHOS ($10^{-7}M_\odot \ltsim M \ltsim 1 M_\odot$)}
\label{GalMACHO}

Massive Compact Halo Objects (MACHOs) were proposed long ago as suitable dark matter candidates constituting the bulk mass of the Milky Way halo. In a most influential paper, Paczy\'nski \cite{1986ApJ...304....1P} proposed a theoretically simple (yet extremely challenging from the observational point of view) method of detecting such compact objects using its gravitational lensing effect. The method consists on observing the temporary increase in brightness produced by the gravitational (micro)lensing effect originated by the occasional alignment of a background star with such compact objects. For the experiment to succeed, it requires a colossal observational effort, consisting in the photometric monitoring of a vast number (e.g. millions) of stars, with high cadence and for very long periods of time. Paczy\'nski showed that the Magellanic Clouds offer the best suitable targets for this purpose. Several independent experiments followed this path, and soon succeeded in detecting the predicted microlensing events, most important being the MACHO \cite{1993Natur.365..621A}, EROS \cite{1993Natur.365..623A} and OGLE \cite{1993AcA....43..289U} collaborations.

The duration of a microlensing event is dependent on the lens mass $M_L$. For the LMC, the rule of thumb (somewhat dependent on the model) is that the duration of a microlensing event (Einstein crossing time) is: $\langle t_E \rangle\sim 70 \sqrt{M_L/M_\odot}$ days. These experiments have been most efficient detecting events with duration between a few days to $\sim$ 100 days and are, therefore, most sensitive to lens masses in the "stellar" range $0.1 - 1 M_\odot$. Indeed, it is in this mass range where most events have been detected and best limits can be set, with limits for larger and lower masses being established based on the absence of excess events from what is expected. The resulting limits are dependent on the halo and galaxy model, importance of self-lensing, etc.

The main results from those microlensing experiments towards the Magellanic Clouds \cite{2000ApJ...542..281A, 2001ApJ...550L.169A, 2007AaA...469..387T, 2011MNRAS.416.2949W}
set abundances below 6-7\% for compact objects in the halo in the mass range between roughtly 0.01 to 0.5 $M_\odot$. The limits relax as we move to higher masses, allowing up to 20\% for objects of a few solar masses. These results are somewhat model dependent, particularly on the mass of the Milky Way halo, but also on the mass and spatial distribution of the lenses \cite{2018MNRAS.479.2889C}. In general, low mass halo models leave some more room for MACHOs in the halo than the high mass models.

Searching for larger mass objects, Blaineau et al. (2022) have recently combined EROS-2 and MACHO data to obtain long time spans (up to 10.6 years) in 14.1 million objects, and searched for long duration events (100-1000 days) founding none\footnote{Two potential candidates remained uncertain and cannot be fully excluded}. This allowed them to establish at 95\% confidence level, an upper limit fraction of 15\% of the halo mass in form of compact objects in the mass range from 10 to 100 $M_\odot$ (again, higher masses are much less restricted by these procedures, allowing up to 50\% for objects of 1000 $M_\odot$). 

Similarly, on the low mass range, Mr\'oz et al. (2017) used OGLE-IV microlensing observations towards the galactic bulge of about 50 million stars with very high cadence ($\sim$ 20 min). From a thorough analysis of the frequency of short events (0.1 to 2 days), they could set stringent limits of a few percent of the mass in compact objects in the planetary mass range from $10^{-5}$ to $10^{-2} M_\odot$. Similar results have been obtained by \cite{2019PhRvD..99h3503N}. Even lower masses have been constrained using the Subaru-HSC survey of 87 million stars in M31 with very high cadence \cite{2019NatAs...3..524N}, even considering finite size effects by \cite{PhysRevD.102.083021}, who constrain abundances below a few percent in the range between $10^{-5}$ and $10^{-10} M_\odot$.  

Overall, galactic microlensing experiments have established quite stringent constraints on the possibility that compact objects contribute significantly to the mass of the Milky Way Halo \cite{2010GReGr..42.2047M}. The results show that it is very unlikely that these compact objects are an important constituent of the halo if they are in the mass range between $10^{-7}$ to $1 M_\odot$. Higher masses are, nevertheless, less restricted, and although compact objects of up to 100 $M_\odot$ are not fully discarded, they cannot be the main halo constituent. 

\section{Extragalactic MACHOS}
\label{ExGalMACHO}

\subsection{Stellar mass range ($0.1 M_\odot \ltsim M \ltsim 3 M_\odot$)}
\label{ExGalMACHO:stel}

On top of the limits set on compact objects in the halo of the Milky Way, gravitational microlensing can also be used to constrain the abundance of such objects in the extragalactic realm.
The idea of observing microlensing  in the images of multiply lensed quasars caused by stars in the lens galaxy was first suggested by \cite{1979Natur.282..561C}. Several authors recognized the possibilities of using quasar microlensing monitoring or single-epoch snapshots of the flux-ratio anomalies induced by microlensing to estimate the optical depth of the microlenses population \cite{1991AJ....102.1939W, 1995ApJ...443...18W, 1996MNRAS.283..225L, 2002ApJ...580..685S}. 
A first practical attempt to implement this idea was carried out by \cite{2004IAUS..220..103S},
using flux-ratio anomalies of a sample of 11 lensed quasars.

A central challenge met by the early extragalactic micro-lensing studies was the determination of a robust base-line of no microlensing magnification. \cite{2009ApJ...706.1451M} solved the problem using the quasar emission lines, arising from a region large enough as to be insensitive to microlensing, to measure the intrinsic flux-ratios between images in a large sample (20) of lensed quasars. The flux-ratio anomalies between images $i$ and $j$ attributable to microlensing are then calculated from,

\begin{equation}
\label{micro}
\Delta m_{ij}=(m_i-m_j)^{cont}-(m_i-m_j)^{line.},
\end{equation}
where we the continuum flux-ratio (expressed as a difference in magnitudes) emitted from a small region of a few light-days in size (according to reverberation mapping studies \cite{2015ApJ...806..129E, 2016ApJ...821...56F, 2017ApJ...836..186J},  and see also the discussion in \cite{2017ApJ...836L..18M}) and hence, strongly affected by microlensing is compared with the intrinsic flux-ratio inferred from the emission lines arising from a large region of several hundred light-days in size (almost insensitive to microlensing). Notice that if the continuum is taken at a wavelength relatively close to the emission line the anomalies estimated in this way are virtually free from extinction effects.

\begin{figure}[ht]
\centering
\includegraphics[scale=.5]{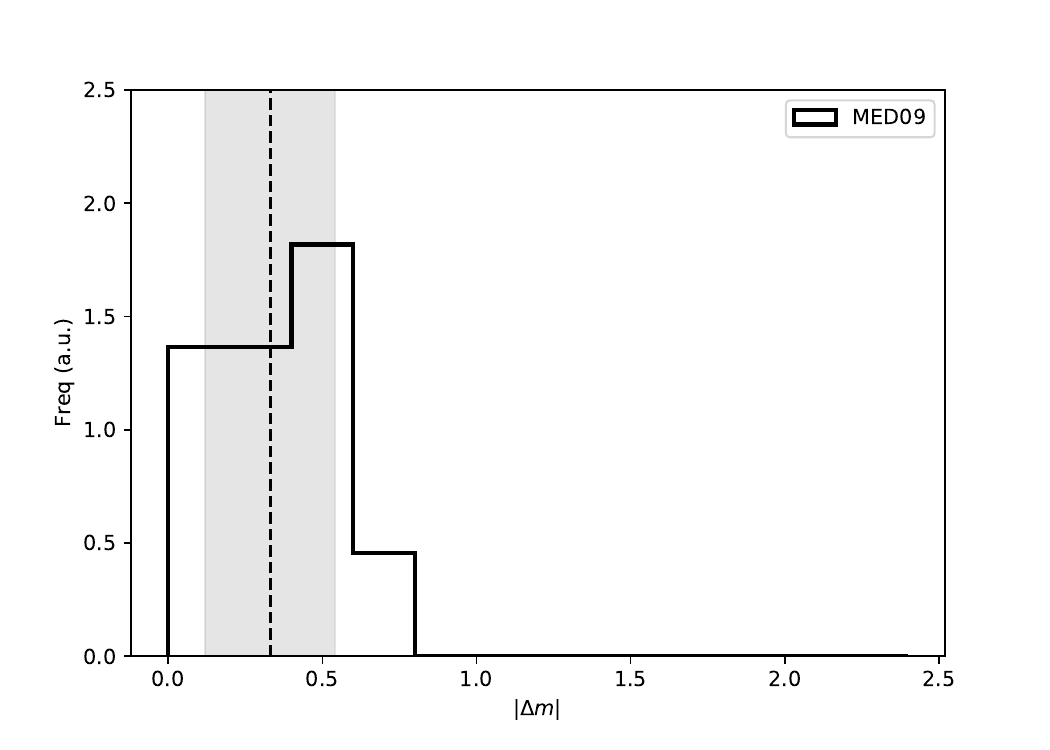}
\caption{.
Histogram of (absolute value of) microlensing magnifications for the UV continuum from a sample of 34 image pair measurements from 23 lensed systems. The vertical dashed line indicates the mean value and the shaded band indicates one sigma around the average.}
\label{fig:F1}       
\end{figure}

\begin{figure}[ht]
\centering
\includegraphics[scale=.5]{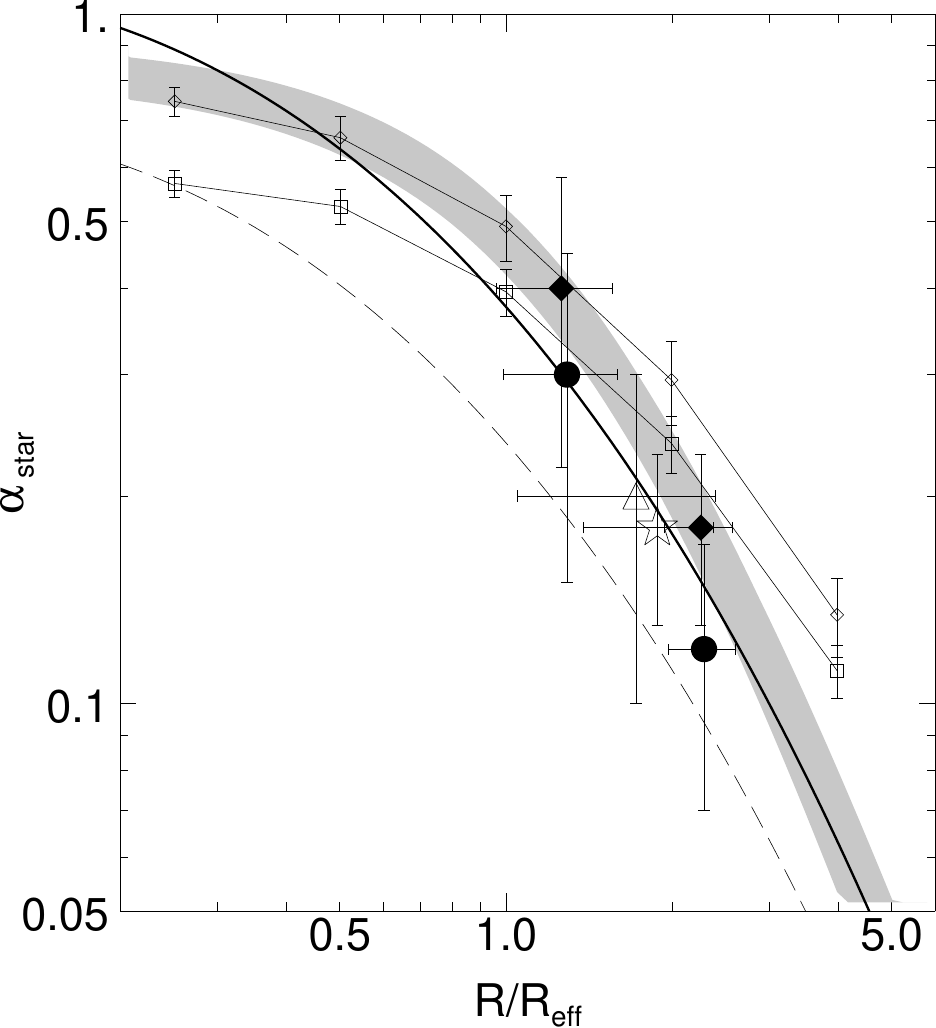}
\caption{Radial profile for the stellar mass fraction. The star (triangle) use only the X-ray (optical) data and a single radial bin for all objects. Circles (diamonds) are the estimates for the two radial bins using a logarithmic (linear) prior on size and both the X-ray and optical data. The dashed line corresponds to a simple model with a de Vaucouleurs stellar component and a total mass corresponding to a SIS with a flat rotation curve equal to the maximum rotational velocity of the stellar component. The thick line is the fiducial galaxy model from \cite{2014ApJ...793...96S}. The gray band is the best fit profile for the sample of lenses analyzed by \cite{2014MNRAS.439.2494O}. Open diamonds and squares correspond to a model using a Hernquist component for the stars, embedded in an NFW halo with (open squares) and without (open diamonds) adiabatic contraction of the dark matter, also from \cite{2014MNRAS.439.2494O}.
Figure adapted from Ref.~\cite{2015ApJ...806..251J}.}
\label{fig:radprof}       
\end{figure}

In Figure \ref{fig:F1} we show an updated histogram of (the absolute value of) microlensing magnifications (see Mediavilla et al. 2024), which basically confirms the results by \cite{2009ApJ...706.1451M}: most of the measurements are concentrated at moderate magnifications, $|\Delta m| \le 0.5\,\rm mag$, and no large magnification tail is present. Next step to estimate the abundance of the compact objects that act like microlenses is to simulate the expected microlensing for different values of the fraction of total mass in microlenses (the total mass is estimated from macro-models). This is usually performed using microlensing magnification maps from which the theoretical microlensing magnification PDFs are obtained. Considering $1M_\odot$ compact objects, \cite{2009ApJ...706.1451M} compare the simulated PDFs with the observed histogram and obtain a maximum likelihood estimate of the mass fraction in  microlenses between $\sim$0.05 and $\sim$0.20 when the source size ranges between $\sim$1 and $\sim$10 light-days. Alternatively, by virtue of the mass-length degeneracy, if we fix the size in 5 light-days (according to reverberation mapping estimates, see above) the results hold in the range between $\sim0.1M_\odot$ and $\sim10M_\odot$. These estimates have been confirmed by posterior studies, jointly analyzing the source size and the mass fraction of the microlenses by \cite{2015ApJ...799..149J}, using X-ray data \cite{2015ApJ...806..251J}, or based on enlarged data samples \cite{2022ApJ...929..123E}.
A summary of the main results about abundance of compact objects in the stellar mass range derived from gravitational lensing, is presented in Figure \ref{fig:radprof}, where we plot the fraction of mass in compact objects versus the radial distance to the center of the galaxy. The triangle corresponds to optical microlensing considering a single radial bin for all the objects from the study of \cite{2015ApJ...799..149J}
. The star corresponds to the average, also in a single radial bin, of the results of a sample of objects observed in X-rays (\cite{2015ApJ...806..251J}). The estimates of the fraction of mass in microlenses based on X-ray microlensing have the advantage that they are insensitive to the size of the source (that in X-ray is significantly smaller than the Einstein radius) for objects in the stellar mass range, cancelling a possible degeneracy. The diamonds are the estimates considering two radial bins for the objects combining both the X-ray and optical data \cite{2015ApJ...806..251J}. The recently reported value of the fraction of mass in stars of 16\% based on microlensing measurements from an enlarged sample of lenses by \cite{2022ApJ...929..123E}  
would also be in reasonable agreement with these results. All the estimates nicely match the fiducial galaxy radial profile (continuous thick line) based on the mass fundamental plane scaling from \cite{2014ApJ...793...96S}  
as well obtained from microlensing observations. Moreover, all these microlensing based results are in good agreement with the radial profile inferred by \cite{2014MNRAS.439.2494O} 
from strong lensing models of a galaxy sample (gray band) and with  a simple theoretical model (de Vaucouleurs + SIS) matching the rotational velocity of the stellar component \cite{2015ApJ...806..251J}. 

Independent limits (albeit less stringent) have been set by \cite{PhysRevLett.121.141101} from gravitational lensing of type Ia supernovae (SNe Ia), who established that compact objects with masses above $0.01 M_\odot$ contribute less than 37\% to the total mass of lens galaxies.

Thus, the main conclusion of the determination of the abundance of stellar mass MACHOS in the extragalactic domain is that quasar microlensing probably arises from the normal star population of lens galaxies and there is no statistical evidence for MACHOS in the dark halos.


\subsection{LIGO-Virgo mass range ($3 M_\odot \ltsim M \ltsim 60 M_\odot$)}
\label{ExGalMACHO:LIGO}

The detection of gravitational waves by the LIGO experiment produced by binary black hole mergers ( \cite{2016PhRvL.116f1102A} 
) involving masses significantly larger than originally expected for black
holes of stellar origin (a typical value of 30$M_\odot$, but with
estimates as large as 60$M_\odot$) renewed the interest in the possibility that dark matter could consist of primordial black holes in the intermediate-mass range, $1 M_\odot \ltsim M \ltsim 1000 M_\odot$ \cite{2016PhRvD..94h3504C}.

\begin{figure}[th]
\centering
\includegraphics[scale=.8]{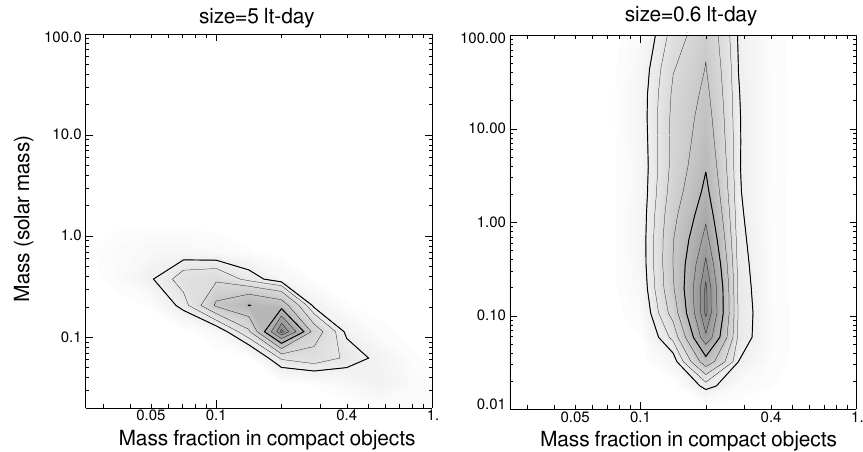}
\caption{Probability distributions for the mass and abundance of compact objects in lens galaxies from optical (left) and X-ray (right) data. Figure adapted from \cite{2017ApJ...836L..18M}.}
\label{fig:F3}       
\end{figure}

To analyze this possibility, \cite{2017ApJ...836L..18M}  extend up to $100M_\odot$ 
the microlensing studies by \cite{2015ApJ...799..149J, 2015ApJ...806..251J} focused on the stellar mass range. To do that, they fix the quasar source size to a fiducial value of 5 light-days in the optical and of 0.6 light-days in X-ray, according to the results from reverberation mapping. Then, they use the mass-length invariance of gravitational lensing to transform the joint probability distributions of the mass fraction of microlenses determined by  \cite{2015ApJ...799..149J, 2015ApJ...806..251J} 
in the optical and in X-rays from the $(\alpha,r)$ plane to the $(\alpha,M)$ plane. The results (see Figure \ref{fig:F3}) strongly constrain the compact object masses to the range $0.05 M_\odot \ltsim M \ltsim 0.45 M_\odot$ at the 90\% confidence level. They also estimate a $\sim$20\% fraction of the total mass in compact objects. Both results are in agreement with the expected masses and abundances of the stellar component. and are incompatible with the existence of any significant population of massive black holes with masses above the typical stellar ones. However the upper value of the mass effectively probed by Mediavilla et al. (2014) is limited by the assumption that the BELs are not affected by microlensing and can be used to define the zero microlensing baseline. If the BLR is a few hundreds of light days large, it may be significantly microlensed by objects of $M_{BH}\gtrsim 100M_\odot$. Thus masses above this upper limit should be explored using other reference to set the no microlensing base-line. 

\begin{figure}[th]
\centering
\includegraphics[scale=.7]{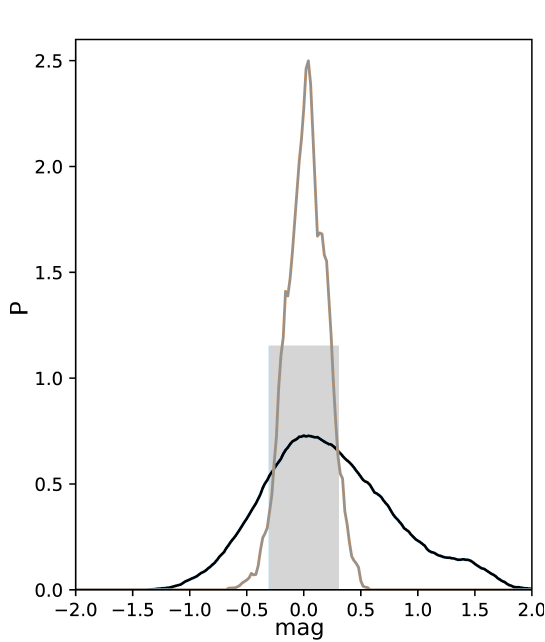}
\caption{Microlensing magnification histograms for a population of 20\% stars with 0.2 $M_\odot$ plus 80\% smooth dark matter (dark grey curve) and a population of 20\% stars plus 80\% LIGO/Virgo type BHs of 30 $M_\odot$ (black curve) for a source of 10 lt-days. The shaded area marks the region containing 70\% of the observed microlensing magnifications by \cite{2007ApJ...661...19P}. Figure adapted from \cite{2022ApJ...929L..17E}.}
\label{fig:Ana}       
\end{figure}

The study by  \cite{2017ApJ...836L..18M} 
is based on the use of a distribution of single-mass microlenses, while the real situation can be a mix of stars and BHs with very different masses. To take into account this,  \cite{2022ApJ...929..123E, 2022ApJ...929L..17E} 
consider a bimodal distribution of stars of $0.2M_\odot$ and BHs of $30M_\odot$. \cite{2022ApJ...929L..17E} 
present a broadly applicable argument based on the sensitivity to the size of the quasar (respect to the Einstein radius defined by the microlens mass) of the probability of observing a given microlensing magnification. They show that there is a strong difference in the predicted microlensing magnifications from a single population of stars and from a mix of stars with LIGO-Virgo BHs when a source of finite size (i.e. of size different from zero) is considered. The Einstein radius of a $\sim30M_\odot$ BH is $\sim 100$light-days, very much larger than the typical size of an accretion disk at the UV ($\sim 5$ light-days, according to reverberation map results), while the Einstein ring of a $0.2M_\odot$ star is less than 10 light-days. Then, quasars approximate the point source limit for the BHs but have a size comparable to the Einstein radius of the stars. In the case of the stars, this result in a smoothing of the magnification changes at the source plane and to a substantial narrowing of the PDF of the predicted magnifications. Comparison with the observed histogram of microlensing magnifications obtained by \cite{2007ApJ...661...19P}, reveals that the predictions for a population of only stars match reasonably well the observed magnifications while a mix of stars and BHs predict too large magnifications (see Figure \ref{fig:Ana}). Consequently, if a significant part of the dark matter of the lens galaxies were in the form of BHs of the masses detected by LIGO/Virgo, much larger microlensing magnifications should have been regularly observed. As this is not the case, a PBHs population can safely be discarded. On the other hand, as the experimental histogram of microlensing magnifications by  \cite{2007ApJ...661...19P} is obtained taking as reference for zero microlensing the flux-ratios between images inferred from macro-models, it is not sensitive to the mass of the microlenses. Thus, taking into account that if a source is point like for a BH it also would be so for any BH of a larger mass, we can discard the existence of a significant population not only for the intermediate mass range but also for any population of BHs of indefinitely large mass.

This result based on a general argument is quantitatively confirmed from a thorough Bayesian statistic study by \cite{2022ApJ...929..123E}. They estimate at once the most likely abundances of stars and BHs, considering an additional smooth dark matter component and a quasar source size in agreement with current reverberation mapping estimates. The Bayesian study is based on the microlensing magnifications of both, the \cite{2009ApJ...706.1451M}  sample extended to include 17 new measurements, and the  \cite{2007ApJ...661...19P} sample. \cite{2022ApJ...929..123E} explore the 10$M_\odot$ to 60$M_\odot$ range for the BH mass, finding an upper limit to the abundance of BH  $\ltsim$0.5\% at the 68\% confidence level. They also estimate a 16\% contribution from the stars, in agreement with previous studies. These results are confirmed when uncertainties in source size and macrolens models are taken into account.

There is another broadly applicable rationale against the existence of a significant population of BHs with masses above $\sim10M_\odot$. \cite{2007ApJ...661...19P}  find that X-ray flux-ratio anomalies are much larger than the optical ones. This result, confirmed by Jimenez-Vicente et al. (2015), indicates that the effect causing the anomalies is sensitive to the differences in size between the X-ray and optical quasar sources, while they should perform as point like under the lensing action of large mass microlenses (\cite{2007ApJ...661...19P}). This is illustrated by Figure 1 of \cite{2022ApJ...929..123E} where we can appreciate the large sensitivity of the microlensing probability distribution corresponding to a $0.2M_\odot$ stellar population to the source size change while the probability associated to a mix of stars and $30M_\odot$ BHs is almost invariable.

\subsection{Substellar mass-range ($2\times 10^{-3} M_\odot \ltsim M \ltsim 0.1 M_\odot$)}
\label{ExGalMACHO:substel}
The effect of microlensing is strongly diluted when the size of the source is much larger than the size of the Einstein radius of the microlenses, as the source {\em feels} a very smoothed version of the magnification structure at the source plane.
Therefore, observations in the visible range (UV rest frame) corresponding to sources with a size of a few light days (which is the size of the Einstein radius of masses in the stellar range), are insensitive to lenses with substellar masses. Very compact sources are therefore needed if we are to probe the substellar mass range of compact objects. With this idea in mind, \cite{2023ApJ...954..172E} have used observations in X-rays, emitted by very compact sources of $\sim1$ lt-day, comparing them with a set of simulated models containing different proportions of stars, smoothly distributed dark matter, and substellar compact objects in the mass range from $\sim$0.1 to $2\times10^{-3} M_\odot$. The mass limits in this study are bounded on the upper side by the confusion with stars (as there is no way microlensing can distinguish the nature of compact objects of similar masses), and on the lower side by the Einstein radius of the lenses compared to the source size (which cannot produce significant magnifications for larger sources). 
The results of this study showed that the fraction of mass in form of stars is pretty well established to $\sim$ 12\% in good agreement with expectations. For the mass in form of compact objects of substellar mass, the estimated values are around 4\%, with upper limits around 10\% for the masses down to 0.01 $M_\odot$. For the lowest studied masses of 0.002 $M_\odot$, whose Einstein radius is similar to the X-ray source sizes, the method is much less restrictive, with an estimated abundance of 10\% (and an upper limit of 34\%) of the total mass in form of this objects. Indeed, the data are fully compatible with 0\% of the mass in the lens in form of objects in this mass range. In conclusion, the majority of the dark matter in these lens galaxies must be in the either in a smooth component or, at least, in the form of objects of smaller mass which cannot produce measurable microlensing efects for the X-ray sources.

Extragalactic single stars would be the optimal sources to probe low mass compact objects in the halos of galaxies via gravitational microlensing given their extremely compact source (compared to the emission from quasars in any band), but even the brightest supergiant stars are, of course, too faint to be observed at large distances. Yet, nature eventually provides unexpected opportunities, and if such star happens to be at the right spot in a galaxy behind a large cluster of galaxies (which provides very large magnifications of a few thousands), it might be magnified enough to be observed at cosmological distances. Since the first claimed observation of a single star at redshift $z=1.5$ in this type of scenario by \cite{2018NatAs...2..334K} as a transient event, a few other examples have been reported, even up to a record distance of $z=6.2$ \cite{2022Natur.603..815W}. The transient nature of these observations in objects which are not intrinsically variable seems to indicate that an extrinsic induced variability is taking place, with gravitational microlensing by compact objects in the cluster halo being the favoured candidate. Several studies \cite{2017ApJ...850...49V,PhysRevD.97.023518} have carefully analyzed the shape of the light curve of such stars for a suitable set of parameters (including the amount of intra-cluster stars, the size of the star and the transverse effective velocity of the source) and have shown the huge potential of observations of these objects to restrict the mass and abundance of compact objects in the halo. In particular,
given the extreme compactness of the source, these observations can probe down to very low masses. Unfortunately, present observations are scarce, and no evidence of a microcaustic crossing has been found. Moreover, there is a large uncertainty with respect to the transverse velocity of the source. Overall, limits on the abundance of objects in the mass range $10^{-5} - 10 M_\odot$ to below 10\% have been established \cite{PhysRevD.97.023518} with this procedure, but there is a large uncertainty in these limits due to the lack of knowledge on the transverse velocity.  There are ongoing observational programs to discover and follow up these lensed stars (some taking benefit of the exquisite resolution and sensitsivity of the JWST). These observations are a promising method to be able to set more stringent limits to compact objects in the low mass range.

\section{Mass Spectrum and Clustering Effects}
\label{massspecandclus}

In the previous sections, we have considered simple mass distributions (single or bimodal) and a reasonably (random) uniform spatial distribution of the microlenses. Nevertheless, formation theories for PBHs do not predict so simple mass and spatial distributions. PBHs should have a continuous mass distribution (as well as stars do) and, under certain circumstances, they could be spatially clustered \cite{Green2021}. It is therefore reasonable to wonder how the previous constraints on the abundance of PBHs from microlensing are affected when these two effects are considered.
Again, there is a significant difference in the approach for galactic (based on the number of events of a certain duration) and extragalactic microlensing (based on flux ratio anomalies or light curves).

A study of galactic microlensing including an extended mass function and spatial clustering has been performed by \cite{2018MNRAS.479.2889C}. The main conclusion of their work is that including an extended mass function and spatial clustering of the PBHs has the effect of shifting the constraints towards smaller masses, therefore leaving some more room for PBH of large mass ($\sim 10 M_\odot$) which become much harder to detect.

In the case of extragalactic studies (mostly from quasar microlensing), we will consider the effect of the extended mass function and the spatial clustering of the lenses separately. 

With respect to the mass function of microlenses, the situation is completely different for single-epoch and light curve studies. While in the latter case, the mass distribution of lenses may, in principle, leave detectable traces in a microlensing light curve, this inclusion would greatly complicate modeling (even more) to an extreme very difficult to handle. Indeed, microlensing is a quite degenerate problem (where mass and abundance of the lenses, size of the source and transverse velocity may combine in several different ways to produce similar observational results). Being able to reproduce the details of observed microlensing light curves while considering a large number of physical parameters is a very difficult problem. This is probably the reason why works dealing with these techniques have considered only a population of stars (for which we have quite a good knowledge from other sources), with none (to our knowledge) considering the contribution from a potential population of PBHs (or other population of compact objects). For the former case of single-epoch microlensing, where observations provide much more limited information, and the overall statistics of magnification is the most significant quantity, the situation is somewhat different, and several works have addressed the issue.

Several studies have shown that for continuous distributions of mass, the specific shape of the mass function is not determinant in the overall microlensing magnification statistics, with the mean mass being the most relevant parameter, except for extremely bimodal distributions with very different characteristic masses (see \cite{2004ApJ...613...77S} and also discussion in \cite{2019ApJ...885...75J}) which may be, indeed, the relevant case for PBHs (mixed with the stellar population of the lens). \cite{2019ApJ...885...75J} already hinted that in this case, the relevant parameter determining the statistics of microlensing magnification of a mix of lenses of different masses was the geometric mean mass instead of the usual arithmetic mean mass\footnote{For most smooth mass distributions, unless extremely wide, the arithmetic mean and the geometric mean mass are indeed very similar.}. Subsequently, \cite{2020ApJ...904..176E} proved that this is also the case for pure bimodal distributions, with special application to the case of PBHs\footnote{The work by \cite{PhysRevD.96.043504} may also open an interesting alternative path to constrain extended mass functions.}.
Therefore, in general, the microlensing effect of a continuous mass distribution, even extremely bimodal, could be modeled by a single mass distribution\footnote{A bimodal continuous distribution can be very well modeled by a two mass distribution where each mass is the characteristic (geometric) mean mass of each population.}. 
With the aim of studying the Initial Mass Function (IMF) of lens galaxies using microlensing, \cite{2019ApJ...885...75J} estimated the average mass of microlenses in lens galaxies to be around 0.16 $M_\odot$, and were able to reproduce this average value with a plausible IMF with a high mass slope and a low mass cutoff
fully compatible with the ones obtained by other means. That implies that the observed microlensing (both in UV rest frame continuum and X-rays) can be fully explained with the stellar population with no need for a population of other compact objects. 
If a populaton of such objects would have contributed significantly to the mass in the lens, they would have been most likely detected, by producing an estimated (geometric) mean mass significantly different of the value expected for a standard stellar population. The only situation that could affect this general statement is if the source size lies in between the Einstein radii characteristic for each of the masses in a bimodal distribution. This scenario of a bimodal mass distribution, taking into account the source size, has been explicitly modelled in the works of \cite{2022ApJ...929..123E, 2022ApJ...929L..17E,2023ApJ...954..172E} whose limits are, therefore, rather general.

Finally, we need to address the issue of spatial clustering of the PBHs, and how this would affect the previous constraints on their abundance. In most studies of (single-epoch) quasar microlensing, the flux ratio anomalies are measured with respect to some baseline which is little or no affected by microlensing (cf. Equation \ref{micro}). This baseline is usually determined by the flux ratio coming from very large sources, like the emission from broad lines (with typical sizes of hundreds of light days), Mid Infrared from the dust torus (with typical size of 1 pc) or radio emission (with a more uncertain size, but of the order of 5 pc or even larger). While these sources are much larger than the Einstein radius for lenses up to $\sim 100 M_\odot$, and are therefore little affected by microlensing by such objects, the situation may be very different if the PBHs are clustered, as in that case, the Einstein radius of the cluster as a whole, can be comparable (or even larger) than those sources which, consequently, can no longer be used as an unmicrolensed baseline.
Heydenreich et al. (priv. comm. 2024) have addressed this issue by using as baseline the flux ratio prediction from the (macro) model
which (although subject to quite a large uncertainty) is virtually free from microlensing effects by the clusters.
They compared flux ratio anomalies in broad lines, MIR and radio (therefore unaffected by microlensing by stars) with numerical simulations of microlensing by clusters of PBHs in the LIGO-Virgo mass range (i.e. $\sim 30 M_\odot$). The modeled clusters have sizes in the range 5 - 15 pc and are composed of 300 to 3000 PBHs according to theoretical predictions \cite{2022JCAP...08..035G}. They set an overall upper limit for clustered PBHs in this mass range of $\sim 3$\% (at 68\% confidence level)\footnote{This figure can increase up to 7\% if uncertainties in the macro lens models are neglected.}, with the highest abundances permitted for the most diluted and poorly populated clusters. Taking into account that diluted and poorly populated clusters approach the limit of uniform distribution, which has been addressed by \cite{2022ApJ...929..123E, 2022ApJ...929L..17E}, we can accept the generality of the overall conclusion: PBHs in the intermediate mass range cannot be a main component of the dark matter in lens galaxies.

\section{Summary of constraints}

\begin{figure}[th]
\centering
\includegraphics[scale=.8]{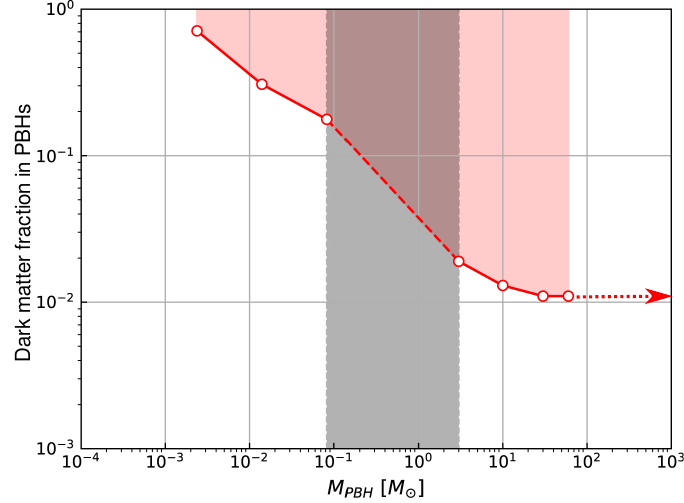}
\caption{Limits on the abundance of PBHs from quasar microlensing observations. The grey band indicates the region of confusion with stars. Limits extend to higher masses. Figure adapted from \cite{2023ApJ...954..172E}.}
\label{fig:F4}       
\end{figure}

As we have seen in Section \ref{GalMACHO}, the galactic microlensing studies can establish rather strong limits ($<$ few percent) for the abundance of PBHs in the halo in the low mass range from $10^-7$ to $\sim 1 M_\odot$. These constraints relax for higher masses ($> 1 M_\odot$), particularly when a mass spectrum and non uniform spatial distributions are taken into account.

In Figure \ref{fig:F4}, we gather the constraints imposed by extragalactic microlensing on the fraction of mass in BHs from the study of a mixed population of BHs, stars and a smooth dark matter component. According to the mass-range, we have considered 4 regions.  (i) $2 \times 10^{-3}M_\odot$ to $10^{-1}M_\odot$. As the lower end of this mass-range is approached, the microlensing magnification signatures attributable to the BHs become progressively washed out (smoothed at the scale of the X-ray source size), and the ability of microlensing to constrain the abundance of BHs diminishes, reaching a 10\% for the smallest studied mass. Above $\sim0.01M_\odot$ a 4\% limit can be established. In any case, note that the abundance  inferred for the stars ($\sim$ 12\%) is consistent with the expectations. (ii) In the second mass-range considered, $10^{-1}M_\odot$ to $3 M_\odot$, BHs and stars become indistinguishable, but also in this case, the consistence of the obtained fraction of mass in compact objects  with the expected abundance of stars ($\sim$ 16\%) support the exclusion of a significant BH population. (iii) In the $3M_\odot$ to $100M_\odot$ range, the upper limit on the abundance of BHs diminishes with the mass from 2\% to 1\%. (iv) As the BH mass increases, the point source limit is reached for the quasar continuum source and we can safely extrapolate the 1\% upper limit to any larger mass, $M>100M_\odot$. This last limit may be important to complement the constraints imposed from other grounds than microlensing in this (unlimited) mass range.

Thus, we can conclude that either from generic arguments or from detailed Bayesian calculations, observations of microlensing magnifications exclude the presence of a significant population of BHs with masses above $10^{-7}M_\odot$. Furthermore, this conclusion holds even when the BHs are distributed following a mass spectrum or appear in clusters.

\label{future}

\section*{Acknowledgement}
We thanks the anonymous referee for valuable comments, useful to improve the paper. This work is financed by grants PID2020-118687GB-C31 and PID2020-118687GB-C31, financed by MCIN/AEI/
10.13039/501100011033. J.J.V. is also financed by projects FQM-108 financed by Junta de Andalucía

\bibliographystyle{unsrt}
\bibliography{main}

\end{document}